\documentclass[11pt,a4paper]{article}
\usepackage{jcappub}
\usepackage[sort&compress]{natbib}
\usepackage{hyperref,ifthen}
\newcommand{\beq}[1]{\begin{equation}\label{#1}}
 \newcommand{\eeq}{\end{equation}}
 \newcommand{\bea}{\begin{eqnarray}}
 \newcommand{\eea}{\end{eqnarray}}
\begin{document}

\title{A new test of $f(R)$ gravity with the cosmological standard rulers in radio quasars}

\author{Tengpeng Xu,$^1$}
\author{Shuo Cao,$^{1\ast}$}
\author{Jingzhao Qi,$^{1}$}
\author{Marek Biesiada,$^{1,2}$}
\author{Xiaogang Zheng,$^1$}
\author{Zong-Hong Zhu$^1$}

\affiliation{ $^1$ Department of Astronomy, Beijing Normal
University, Beijing 100875, China; \emph{caoshuo@bnu.edu.cn} \\
$^2$ Department of Astrophysics and Cosmology, Institute of Physics,
University of Silesia, Uniwersytecka 4, 40-007 Katowice, Poland }

\abstract{ As an important candidate gravity theory alternative to
dark energy, a class of $f(R)$ modified gravity, which introduces a
perturbation of the Ricci scalar $R$ in the Einstein-Hilbert action,
has been extensively applied to cosmology to explain the
acceleration of the universe. In this paper, we focus on the
recently-released VLBI observations of the compact structure in
intermediate-luminosity quasars combined with the
angular-diameter-distance measurements from galaxy clusters, which
consists of 145 data points performing as individual cosmological
standard rulers in the redshift range $0.023\le z\le 2.80$, to
investigate observational constraints on two viable models in $f(R)$
theories within the Palatini formalism: $f_1(R)=R-\frac{a}{R^b}$ and
$f_2(R)=R-\frac{aR}{R+ab}$. We also combine the individual standard
ruler data with the observations of CMB and BAO, which provides
stringent constraints. Furthermore, two model diagnostics, $Om(z)$
and statefinder, are also applied to distinguish the two $f(R)$
models and $\Lambda$CDM model. Our results show that (1) The quasars
sample performs very well to place constraints on the two $f(R)$
cosmologies, which indicates its potential to act as a powerful
complementary probe to other cosmological standard rulers. (2) The
$\Lambda$CDM model, which corresponds to $b=0$ in the two $f(R)$
cosmologies is still included within $1\sigma$ range. However, there
still exists some possibility that $\Lambda$CDM may not the best
cosmological model preferred by the current high-redshift
observations. (3) Given the current standard ruler data, the
information criteria indicate that the cosmological constant model
is still the best one, while the $f_1(R)$ model gets the smallest
observational support. (4) The $f_2(R)$ model, which evolves quite
different from $f_1(R)$ model at early times, still significantly
deviates from both $f_1(R)$ and $\Lambda$CDM model at the present
time.}

\maketitle

\section{Introduction}

In order to explain the accelerated expansion of the Universe, which
was strongly supported by the observations of the type Ia supernovae
(SN Ia) \cite{SN1998}, a mysterious component with negative pressure
was introduced as a new cosmological component dubbed as dark energy
(DE). Following this this direction, based on the cosmological
principles (homogeneous, isotropic) and Einstein's general
relativity (GR), the current standard cosmological model considers
the cosmological constant ($\Lambda$) corresponding to a
modification of the energy-momentum tensor in Einstein equations,
which is generally consistent with most of the observational data
including SN Ia, the cosmic microwave background (CMB) \cite{CMB},
the baryon acoustic oscillations (BAO) \cite{BAO}, etc. However,
considering the fact that $\Lambda$CDM is still confronted with the
well-known coincidence problem and fine-tuning problem, a large
number of dark energy models have been proposed to explain the
cosmic acceleration \cite{Chaplygin_gas,quintessence,phantom}. On
the other hand, without introducing the new component in the
universe, the modification of general relativity provides another
way to go. Some popular theories of modifying GR include $f(R)$
theory of gravity \cite{f(R)_theories_1,f(R)_theories_2}, $f(T)$
gravity \cite{bengochea2009dark,cai2015f,qi2016transient},
Gauss-Bonnet and $f(G)$ gravity \cite{nojiri2005modified}. In this
paper, we focus on the so-called $f(R)$ gravity, the advantage of
which not only lies in its ability to explain the late-time cosmic
acceleration of the universe, but also describe the large scale
structure distribution of the Universe
\cite{dos2016n,voivodic2017modeling}. In the framework of $f(R)$
theories of gravity, the Einstein-Hilbert Lagrangian can be modified
by changing the Ricci scalar $R$ to $f(R)$, a differentiable
function of $R$, on the base of which the generalized field equation
is derived by varying the action with respect to the metric.
However, since this method always leads to fourth-order equations
along with some instability problems in many interesting cases
\cite{instability_problems}, we consider in our analysis a different
method called Palatini approach
\cite{f(R)_theories_1,f(R)_theories_2}, which takes the metric and
the connection as two independent field variables in the action, and
varies the action respectively with the two variables to obtain the
generalized Einstein field equation.

From the observational point of view, it is also very important to
explore such $f(R)$ cosmological models in light of observational
data, which has been broadly studied in the literature
\cite{carvalho2008cosmological,de2016theoretical,song2007cosmological}.
There are two direct standard probes of expansion history of the
Universe, one is the standard candles providing the information of
luminosity distance ($D_L(z)$), and the other is the standard rulers
related to the so-called angular diameter distance ($D_A(z)$). For
instance, BAO and CMB peak locations are respectively recognized as
the standard rulers \cite{BAO_1,BAO_2,BAO_3}, and the increasing
observations of these two distance indicators have been widely used
in various cosmological studies. Other most commonly used standard
rulers in cosmology also include the strongly gravitationally lensed
systems
\cite{lensing_1,cao2011constraints,cao2012testing,cao2012SL,cao2015SL},
the x-ray gas mass fraction of galaxy clusters \cite{cao2014cosmic},
etc. In this paper we investigate the constraints on the $f(R)$
gravity from the latest measurement of angle diameter distance from
the recently-released VLBI observations of the compact structure in
intermediate-luminosity quasars \cite{CAO,cao2016cosmology} combined
with the angular-diameter-distance measurements from galaxy clusters
\cite{CLUSTERS}, which consists of 145 data points performing as
individual cosmological standard rulers in the redshift range
$0.023\le z\le 2.80$. Compared with other standard rulers
extensively used in the literature (BAO, strong lensing systems),
quasars (QSO) are observed at much higher redshits ($z \sim 3.0$),
which indicates their potential to test the $f(R)$ theory with the
newly revised observations. Previous papers have demonstrated the
success of this sample in its cosmological application. In Paper I
\citep{CAO}, we demonstrated the existence of dark energy in the
Universe with high significance and estimated the speed of light
referring to a distant past ($z=1.70$). In Paper II and III
\cite{cao2016cosmology,qi2017ft}, we investigated the cosmological
application of this data set and obtained stringent constraints on
the parameters in various dark energy models and $f(T)$ gravity
models. More specifically, in this paper we try to give a new
approach to constraining two viable $f(R)$ models within the
Palatini formalism: $f_1(R)=R-\frac{a}{R^b}$ and
$f_2(R)=R-\frac{aR}{R+ab}$, both of which can result in the
radiation-dominated, matter-dominated and recent accelerating state.
In order to discuss the differences between the two $f(R)$ models
and $\Lambda$CDM model, we apply the information criteria (IC)
\citep{AIC,BIC}, the $Om(z)$ diagnostic \cite{Om_diagnostic} and the
statefinder diagnostic \cite{SF} to analyse different cosmological
models.

This paper is organized as follows. In Section 2, we briefly
describe the basic theory of $f(R)$ gravity as well as the
corresponding cosmological models. In Section 3, we describe the
methodology and observational samples for angular diameter
distances. In Section 4, we perform a Markov Chain Monte Carlo
(MCMC) analysis, and furthermore apply model diagnostics in Section
5. Finally, the conclusions are summarized in Section 6. The units
with constant speed of light $c = 1$ is used throughout this work.

\section{The $f(R)$ gravity in the metric formalism}

In this section we firstly derive the modified Einstein field
equations in $f(R)$ theory as well as the new Friedman equation in
$f(R)$ cosmology, and then briefly introduce the two $f(R)$ models
in Palatini formalism to be considered in this paper.

In Palatini formalism, the modified Einstein-Hilbert action in the
framework of $f(R)$ gravity is given by
\begin{equation}
\label{actionPL} S = \frac{1}{2\kappa}\int d^4x\sqrt{-g}f(R) +
S_{\rm{M}}(g_{\mu\nu},\psi)\,,
\end{equation}
where $\kappa = 8\pi G$, $g$ is the determinant of the metric
$g_{\mu\nu}$, $f$ is a differentiable function of the Ricci scalar
$R$, and $S_M$ is the action of matter depending on the matter field
$\psi$ and the metric $g_{\mu\nu}$. It should be noted that the
metric $g_{\mu\nu}$ and the affine connection $\hat{\Gamma}^\rho
_{\mu\nu}$ are treated as two independent fields for the gravity
action in Palatini formalism. From the affine connection
$\hat{\Gamma}^\rho _{\mu\nu}$, the expression of the generalized
Ricci tensor can be written as
\begin{equation}
\label{ricci_ts} \hat{R}_{\mu\nu} = \hat{\Gamma}^\sigma _{\ \mu\nu
,\sigma} - \hat{\Gamma}^\sigma _{\ \mu\sigma ,\nu} +
\hat{\Gamma}^\sigma _{\ \sigma\rho}\hat{\Gamma}^\rho _{\ \mu\nu} -
\hat{\Gamma}^\sigma _{\ \mu\rho}\hat{\Gamma}^\rho _{\ \sigma\nu}\,,
\end{equation}
and the Ricci scalar is dependent on the metric and the affine
connection as $R = g^{\mu\nu}\hat{R}_{\mu\nu}$. Varying the action
(\ref{actionPL}) with respect to the connection $\hat{\Gamma}$
yields the equation
\begin{equation}
\hat{\nabla}_\alpha[f^\prime(R)\sqrt{-g}g^{\mu\nu}] = 0\,,
\end{equation}
where $f^\prime(R) \equiv \mathrm{d}f/\mathrm{d}x$, and
$\hat{\nabla}$ is the covariant derivative corresponding to the
affine connection $\hat{\Gamma}$. From this equation it can be
easily found that, considering a new metric $h_{\mu\nu} =
f^\prime(R)g_{\mu\nu}$ conformal with the original metric
$g_{\mu\nu}$, the affine connection $\hat{\Gamma}$ can be written as
the usual Levi-Civita connection of the new metric $h_{\mu\nu}$
\begin{equation}
\hat{\Gamma}^\lambda _{\ \mu\nu} =
\frac{h^{\lambda\sigma}}{2}(h_{\nu\sigma,\mu} + h_{\mu\sigma,\nu} -
h_{\mu\nu,\sigma}).
\end{equation}
Thus we can derive the relation between the generalized affine
connection $\hat{\Gamma}$ and the metric $g_{\mu\nu}$ as
\begin{equation}
\hat{\Gamma}^\lambda _{\ \mu\nu} = \Gamma^\lambda _{\ \mu\nu} +
\frac{1}{2f^\prime}[2\delta^\lambda _{\ (\mu}\partial _{\nu
)}f^\prime - g^{\lambda\tau}g_{\mu\nu}\partial _\tau f^\prime].
\end{equation}
Substituting the above equations into Eq.~(\ref{ricci_ts}), the
dependency between the generalized Ricci tensor $\hat{R}_{\mu\nu}$
and the original Ricci tensor $R_{\mu\nu}$ can be expressed as
\begin{equation}
\hat{R}_{\mu\nu} = R_{\mu\nu} - \frac{3}{2} \frac{\nabla_\mu
f^\prime \nabla_\nu f^\prime}{f^{\prime 2}} +
\frac{\nabla_\mu\nabla_\nu f^\prime}{f^\prime} +
\frac{1}{2}g_{\mu\nu}\frac{\nabla^\mu\nabla_\nu f^\prime}{f^\prime}.
\end{equation}

From the discussion above, one can see that varying the action
(\ref{actionPL}) with respect to the connection $\hat{\Gamma}$ gives
the new geometric properties of space-time with the metric
$g_{\mu\nu}$. On the other hand, by varying the action
(\ref{actionPL}) with the metric $g_{\mu\nu}$, we can get the
generalized Einstein field equation
\begin{equation}
\label{field_eq} f^\prime\hat{R}_{\mu\nu} - \frac{f}{2}g_{\mu\nu} =
\kappa T_{\mu\nu}\,,
\end{equation}
where $T_{\mu\nu}$ is the energy-momentum tensor of matter. It is
obvious that when $f(R)=R$, Eq.~(\ref{field_eq}) will reduce to the
Einstein field equation, while Eq.~(\ref{field_eq}) recovers the
Einstein field equation in $\Lambda$CDM model when
$f(R)=R-2\Lambda$.

Now we will introduce the $f(R)$ cosmology in the framework of flat
Friedman - Robertson - Walker (FRW) metric given by
\begin{equation}
\mathrm{d}s^2 = -\mathrm{d}t^2 + a^2 (t)(\mathrm{d}x^2 +
\mathrm{d}y^2 + \mathrm{d}z^2 )\,
\end{equation}
where $a(t)$ is the scale factor related to the redshift $z$ as:
$a(t)=(1+z)^{-1}$. Then the Hubble parameter can be expressed in
terms of the scale factor: $H=\dot{a}/a$, where the overdot denotes
the derivative with respect to the cosmic time $t$. Here we adopt
the Hubble constant prior $H_0=69.6\pm 0.7~ \rm{km} \rm{s}^{-1}
\rm{Mpc}^{-1}$ from the recent measurements of Hubble constant with
1\% uncertainty \cite{bennett2014the}. In the $f(R)$ cosmology,
considering the cosmic fluid as a pressureless dust satisfying
$p_m=0$, we have $T_{\mu\nu} = (\rho_{\rm m}+p_m)U_\mu U_\nu +
p_{\rm m} g_{\mu\nu} = \rho_{\rm m} U_\mu U_\nu$, where $U_\mu$,
$\rho_{\rm m}$ and $p_{\rm m}$ are the $4-$velocity of the fluid,
the energy density and the fluid pressure, respectively. The trace
of $T_{\mu\nu}$ is $T = g^{\mu\nu}T_{\mu\nu} = -\rho_{\rm m} =
-\rho_{\rm m0}(1+z)^3$. Contracting Eq.~(\ref{field_eq}) with
$g^{\mu\nu}$, we can derive
\begin{equation}
\label{field_tr_eq} Rf^\prime(R)-2f(R) = -\kappa\rho_{\rm m} =
-\kappa\rho_{\rm m0}(1+z)^3.
\end{equation}
Taking the time derivative of Eq.~(\ref{field_tr_eq}) and combining
it with the conservation equation $\dot{\rho}_{\rm m} +3H\rho_{\rm
m}=0$, we obtain
\begin{equation}
\label{R_dot} \dot{R} = \frac{3H\kappa\rho_{\rm
m}}{Rf^{\prime\prime}-f^\prime}.
\end{equation}
Next we can derive the generalized Friedman equation by using the
generalized Ricci tensor
\begin{equation}
\label{Friedman_eq} 6(H + \frac{1}{2}
\frac{\dot{f}^\prime}{f^\prime})^2 = \frac{ 3f - Rf^\prime
}{f^\prime}.
\end{equation}
Substituting Eq.~(\ref{R_dot}) into Eq.~(\ref{Friedman_eq}), the
Friedman equation in $f(R)$ cosmology expresses as
\begin{equation}
\label{Friedman_eq1} H^2 = \frac{ 3f - Rf^\prime }{6f^\prime\eta^2},
\end{equation}
where
\begin{equation}
\eta = 1-
\frac{3}{2}\frac{f^{\prime\prime}}{f^\prime}\frac{Rf^\prime
-2f}{Rf^{\prime\prime}-f^\prime}.
\end{equation}
Correspondingly, the angular diameter distance at redshift $z$ in
flat FRW metric reads
\begin{eqnarray}
\label{D_A} D_{\rm A}(z)&=&\frac{1}{1+z}\int _0 ^z
\frac{\mathrm{d}z}{H(z)}
\\ \nonumber
&=&\frac{1}{3}(2f-Rf^\prime)^{-\frac{1}{3}}\int^{R_z}_{R_0}{\frac{Rf^{\prime\prime}-f^\prime}{(2f-Rf^\prime)^\frac{2}{3}}\frac{dR}{H(R)}}
\end{eqnarray}

In this paper, we consider two specific viable $f(R)$ models in the
Palatini approach
\begin{equation}
\begin{split}
f_1(R) &=R-\frac{a}{R^b},\\ f_2(R) &=R-\frac{a}{1+\frac{ab}{R}}.
\nonumber
\end{split}
\end{equation}
As the most commonly used parametrization in $f(R)$, the first model
has been tested with various observational data in many previous
works \cite{f1_R-1,f1_R-2,f1_R-3,f1_R-4,f1_R-5,f1_R-6}. The second
model is originated from the well-known Hu $\&$ Sawicki model
\cite{f2_R}, $f(R)=R-m^2\frac{c_1(R/m^2)^n}{1+c_2(R/m^2)^n}$.
Considering that $n$ is an integer satisfying $n>0.9$, for
simplicity we have set $n = 1$ and thus $a=m^2c_1/c_2$ and $b=1/c_1$
in this work. In the two $f(R)$ models, one can easily find that
\begin{equation}
\lim_{b \to 0} f(R)=R-a,
\end{equation}
which is equivalent to the $\Lambda$CDM model: $f(R)=R-2\Lambda$,
where $\Lambda$ is the cosmological constant. Therefore, there are
two independent model parameters ($a, b$) in the two $f(R)$ models,
both of which can reduce to $\Lambda$CDM when $b \to 0$. We note
that $b$ can be regarded as the deviation parameter quantifying the
deviation of $f(R)$ gravity from $\Lambda$CDM. Moreover, for a
specific $f(R)$ model with certain value of $a$ and $b$, the
combination of Eq.~(\ref{Friedman_eq1}) and (\ref{field_tr_eq}) will
provide the matter density parameter $\Omega_{\rm m}$, which is
defined as $\Omega_{\rm m}= \kappa\rho_{\rm m0}/(3H^2 _0)$.
Therefore, constraint results on the $f(R)$ models can be shown in
the ($a, b$) or ($\Omega_{\rm m}, b$) plane, which will be
specifically presented in section 4.

\section{Observational data and fitting method}

\subsection{Quasars and galaxy clusters}

It has been a long-time controversy that whether compact structure
in radio sources could provide a new type of standard ruler in the
universe
\cite{standard_rulers_1,standard_rulers_2,standard_rulers_3,cao2015exploring}.
Recently, based on a 2.29 GHz VLBI all-sky survey of 613
milliarcsecond ultra-compact radio sources
\cite{compact_radio_1,compact_radio_2}, \citet{CAO} presented a
method to divide the full sample into different sub-samples,
according to their optical counterparts and intrinsic luminosity.
The final results indicated that intermediate-luminosity quasars
show negligible dependence on both redshifts $z$ and intrinsic
luminosity $L$, and thus represent a fixed comoving-length of
standard ruler. More recently, based on a
cosmological-model-independent method to calibrate the linear sizes
$l_m$ of intermediate-luminosity quasars, Cao et al.
\cite{cao2016cosmology} investigated the cosmological application of
this data set and obtained stringent constraints on both the matter
density parameter $\Omega_{\rm m}$ and the Hubble constant $H_0$,
which agree well with the recent \textit{Planck} results. The
constraining power of the quasar data was also studied in viable
$f(T)$ gravity models, where $T$ is the torsion scalar in
teleparallel gravity \cite{qi2017ft}.

In our analysis, we will use the observations of 120
intermediate-level quasars covering the redshift range $0.46 < z <
2.80$, while the linear size of this standard ruler is calibrated to
$l_m=11.03\pm0.25\ {\rm pc}$ by a new cosmological-independent
technique (see \citet{cao2016cosmology} for details and reference to
the source papers). The observable in this data set is the angular
size of the compact structure in radio quasars, the theoretical
counterpart of which expresses as
\begin{equation}
\theta(z)=\frac{l_m}{D_{\rm A}(z)}
\end{equation}
where $D_A$ is the angular diameter distance at redshift $z$
(Eq.(\ref{D_A})). In this work minimize the $\chi^2$ objective
function to derive model parameters of the $f(R)$ theory:
\begin{equation}
\chi^2 _{\rm QSO}(z;\textbf{p})=\sum_{i=1}^{120}\frac{[\theta_{\rm
th}(z_i;\textbf{p})-\theta_{\rm obs}(z_i)]^2}{\sigma_\theta(z_i)^2},
\end{equation}
where \textbf{p} represents the model parameters in $f(R)$ gravity,
and $\theta_{\rm th}(z_i;\textbf{p})$ is the theoretical value of
the angular size at redshift $z_i$, while $\theta_{\rm obs}(z_i)$
and $\sigma_{\theta}(z_i)$ are the observed value and the
corresponding 1$\sigma$ uncertainty of angular size for each quasar,
respectively. Moreover, we have considered the intrinsic spread in
linear sizes $l_m$ by adding $10\%$ systematical uncertainty in the
observed angular sizes in computing. One should note that, in the
framework of $\chi^2$ minimization method, the additional 10\%
uncertainties in the observed angular sizes is equivalent to adding
an additional 10\% uncertainty in the linear size, although the
best-fit parameters describing the dependence of $l_m$ on the
luminosity and redshift are negligibly small
\citep{CAO,cao2016cosmology}.

Moreover, the observations from Sunyaev-Zeldovich effect (SZE) and
X-ray surface brightness from galaxy clusters also offer a source of
angular diameter distances. Considering the redshift range of quasar
sample used in our analysis, i.e., the lack of low-redshift quasars
at $z<0.5$, we also consider 25 galaxy clusters covering the
redshift range $0.023 \le z \le 0.784$ from De Filippis et al.
(2005) sample \cite{clusters1}, in which a isothermal elliptical
$\beta$ model was used to describe the clusters by combining their
SZE and X-ray surface brightness observations. As previously noted
by Ref.~\cite{cao2016cosmology}, due to the the redshift coverage of
high-redshift quasars and low-redshift clusters, the combination of
these two astrophysical probes could contribute a relatively
complete source of angular diameter distances. The observable in
this data is the angular diameter distance $D_{\rm A,obs}$, the
theoretical counterpart $D_{\rm A,th}$ is defined in
Eq.~(\ref{D_A}). Similar to the quasars data, the $\chi^2$ function
for the galaxy cluster data is given by
\begin{equation}
\chi^2 _{\rm cluster}(z;\textbf{p})=\sum_{i=1}^{25}\frac{[D_{\rm
A,th}(z_i;\textbf{p})-D_{\rm A,obs}(z_i)]^2}{\sigma_{D_{\rm
A}}(z_i)^2}.
\end{equation}
where $\sigma_{D_{\rm A}}$ is the 1$\sigma$ uncertainty of the
angular diameter distance for each galaxy cluster.

\subsection{BAO and CMB data}

Besides the individual standard rulers, observations the other two
standard rulers that we shall use in this paper for the joint
cosmological analysis are the BAO and CMB data.

For CMB, we use the measurement of the shift parameter
$\mathcal{R}$, which is sensitive to the distance to the decoupling
epoch corresponding to the overall amplitude of the acoustic peaks.
In $f(R)$ cosmology it can be expressed as
\begin{eqnarray}
\mathcal{R} &=& \sqrt{\Omega_{\rm m}H^2 _0}\int_0 ^{z_\ast}
\frac{\mathrm{d}z}{H(z)} \\ \nonumber &=& 3^{-4/3}(\Omega_{\rm
    m}H^2 _0)^{1/6}\int_{R_0} ^{R_{*}}
\frac{Rf^{\prime\prime}-f^\prime}{(2f-Rf^\prime)^{2/3}}\frac{\mathrm{d}R}{H(R)},
\end{eqnarray}
where the redshift of photon-decoupling period can be calculated as
\citep{Hu96}
\begin{equation}
z_\ast=1048[1+0.00124(\Omega_{b}h^2)^{-0.738}][1+g_1(\Omega_{m}h^2)^{g_2}]
\end{equation}
\begin{equation}
g_1=\frac{0.0783(\Omega_{b}h^2)^{-0.238}}{1+39.5(\Omega_{b}h^2)^{0.763}},
g_2=\frac{0.560}{1+21.1(\Omega_{b}h^2)^{1.81}}.
\end{equation}
In this analysis, the baryon density is fixed at
$\Omega_{b}h^2=0.02222$ and the shift parameter is taken as
$\mathcal{R}=1.7499\pm 0.0088$ from the first year data release of
\emph{Planck} observations \cite{CMB_DATA}. Therefore the $\chi^2$
can be defined as
\begin{equation}
\chi^2 _{\rm CMB} = \frac{(\mathcal{R}-1.7499)^2}{0.0088^2}.
\end{equation}

For BAO, we turn to the latest observations of acoustic-scale
distances from the 6-degree Field Galaxy Survey (6dFGS) at lower
redshift $z=0.106$ \citep{Beutler11}, the Sloan Digital Sky Survey
(SDSS-DR7) catalog combined with galaxies from 2dFGRS (at effective
redshift $z=0.2$ and $z=0.3$) \citep{Percival10}, while the
higher-$z$ measurement is derived from the WiggleZ galaxy survey,
which reported distances in three correlated redshift bins between
0.44 and 0.73 \citep{Blake12}. More specifically, we use the
measurement of distance ratio
$\mathcal{A}(z_{BAO})=d_A(z_\ast)/D_V(z_{BAO})$ from the BAO peak to
set constraint on $f(R)$ model parameters, where $D_{\rm V}$ is the
dilation scale and $d_{\rm A}$ is the co-moving angular diameter
distance (different from the angular diameter distance $D_{\rm A}$).
The expressions of the two types of distances respectively read
\begin{eqnarray}
d_{\rm A}(z) &=& \int^z _0\frac{\mathrm{d}z^\prime}{H(z^\prime)}=(1+z)D_{\rm A}(z) \\
D_{\rm V}(z) &=& (d_{\rm A}(z)^2\frac{z}{H(z)})^{1/3}.
\end{eqnarray}
Observations of the distance ratio at six different $z_{\rm BAO}$
are summarized in Ref.~\cite{BAO_DATA} and explicitly shown in Table
I. Now we can define the $\chi^2$ for BAO as
\begin{equation}
\chi^2 _{\rm BAO} = \sum_{ij}X_i C^{-1} _{ij} X_j,
\end{equation}
with the difference between the theoretical and observational
distance priors $X_i=\mathcal{A}_{\rm th}(z_i)-\mathcal{A}_{\rm
obs}(z_i), i=1,2,...,6$ and the inverse covariance matrix shown in
Table II \cite{BAO_DATA}.

\begin{table*}
    \begin{center}
        \begin{tabular}{|c|c|c|c|c|c|c|}
            \hline \hline
            \small{$z_{BAO}$} & \footnotesize{$0.106$} & \footnotesize{$0.20$}& \footnotesize{$0.35$}&\footnotesize{$0.44$}&\footnotesize{$0.60$} & \footnotesize{$0.73$} \\
            \hline
            \small{Survey} & \footnotesize{6dFGS} & \footnotesize{SDSS-DR7}& \footnotesize{SDSS-DR7}&\footnotesize{WiggleZ} &\footnotesize{WiggleZ} & \footnotesize{WiggleZ} \\
            \hline
            \small{$\frac{d_A(z_\ast)}{D_V(z_{BAO})}$} &\footnotesize{$\ 30.95\pm1.46\ $ }& \footnotesize{$\ 17.55\pm0.60\ $ }&\footnotesize{$\ 10.11\pm0.37\ $ }& \footnotesize{$\ 8.44\pm0.67\ $}&\footnotesize{ $\ 6.69\pm0.33\ $}& \footnotesize{\ $5.45\pm0.31\ $}\\
            \hline \hline
        \end{tabular}
        \caption{ The measurement of distance ratio $d_A(z_*)/D_V(z_{BAO})$ from the BAO observations
        \cite{Beutler11,Percival10,Blake12}.}
    \end{center}
\end{table*}

\begin{table*}[!htpp]
    \centering
    \begin{tabular}{|c|c|c|c|c|c|c|}
        \hline
        $C^{-1}_{ij}$ &1&2&3&4&5&6\\
        \hline
        1&0.48435 & -0.101383 &-0.164945 &-0.0305703 &-0.097874 & -0.106738\\
        \hline
        2&-0.101383 & 3.2882 & -2.45497 & -0.0787898 & -0.252254 & -0.2751\\
        \hline
        3&-0.164945 & -2.45497 & 9.55916 & -0.128187 & -0.410404 & -0.447574\\
        \hline
        4&-0.0305703 & -0.0787898 & -0.128187 & 2.78728 & -2.75632 & 1.16437\\
        \hline
        5&-0.097874 & -0.252254 & -0.410404 & -2.75632 & 14.9245 & -7.32441 \\
        \hline
        6&-0.106738 & -0.2751 & -0.447574 & 1.16437 & -7.32441 & 14.5022  \\
        \hline
    \end{tabular}
    \caption{ The inverse covariance matrix of the BAO observations \cite{BAO_DATA}.}
\end{table*}

\begin{figure*}
\centering
\includegraphics[width=8cm,height=7.2cm]{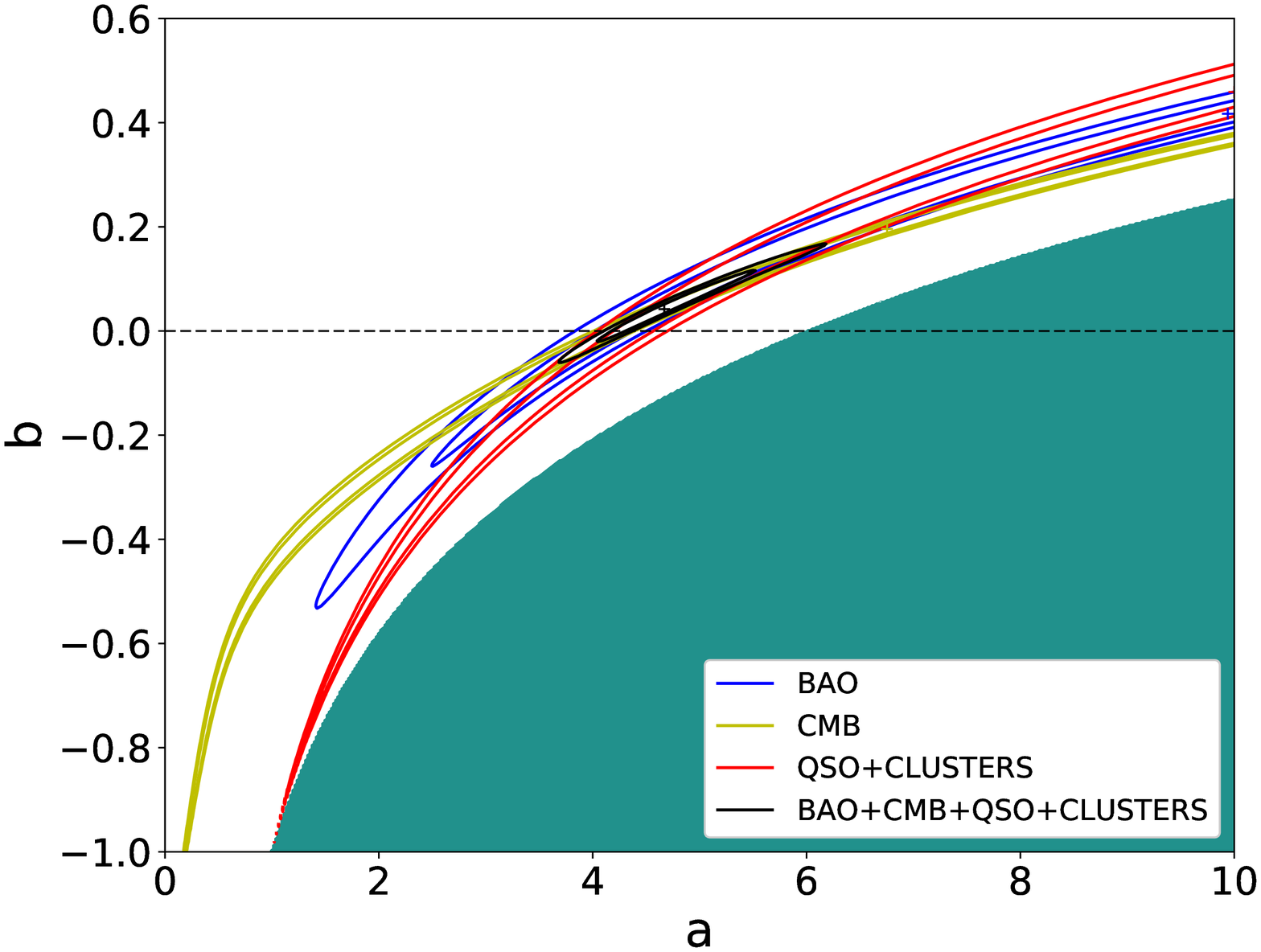}\includegraphics[width=8cm,height=7.2cm]{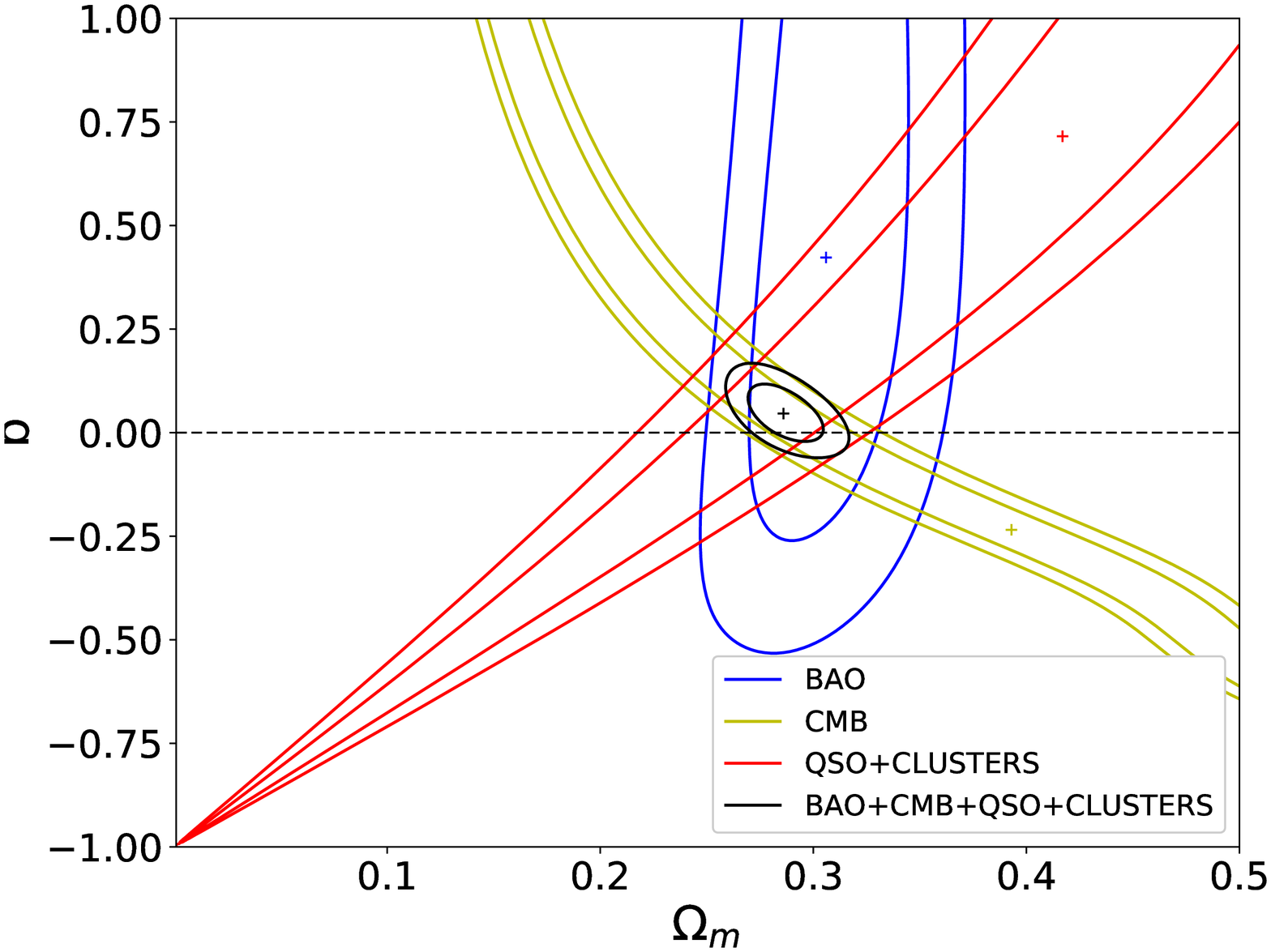}
\caption{ Confidence intervals at $68.3\%$ and $95.4\%$ on the ($a,
b$) and ($\Omega_{\rm m}, b$) planes for the $f_1(R)$ model, arising
from the combined fit including quasars, galaxy clusters, BAO and
CMB data. The gray area represents a section of the parameter space
that is not allowed, dashed line indicates the $\Lambda$CDM model
with $b=0$. }\label{figh}
\end{figure*}

\begin{figure*}
\centering
\includegraphics[width=8cm,height=7.2cm]{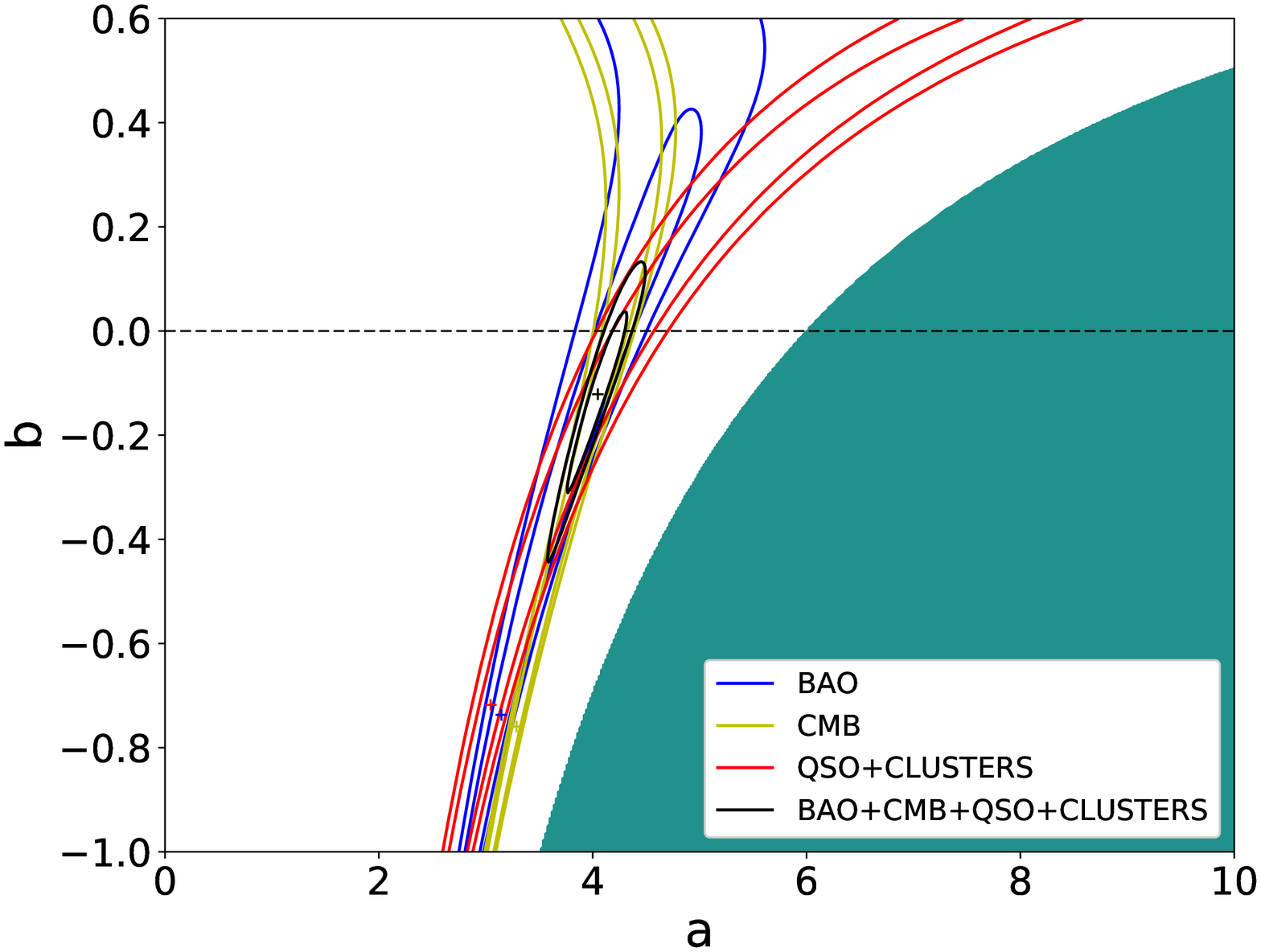}\includegraphics[width=8cm,height=7.2cm]{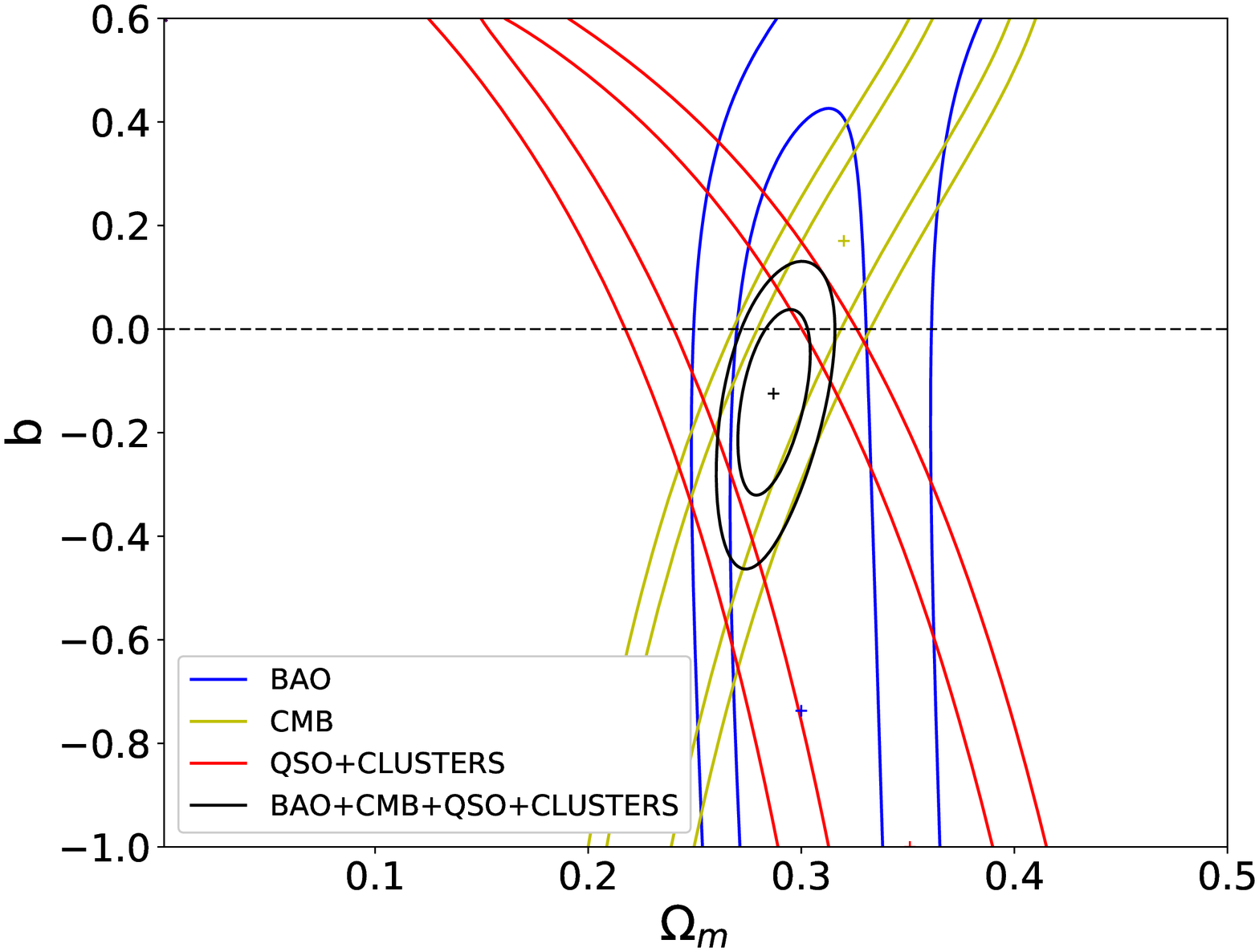}
\caption{The same as Fig.~1, but for the $f_2(R)$
model.}\label{figh1}
\end{figure*}

\section{Observational constraints}

In this section, we present the constraint results of two $f(R)$
models by using different angular diameter distance data,
QSO+Cluster (individual standard rulers), BAO+CMB (statistical
standard rulers), and QSO+Cluster+BAO+CMB (combined standard
rulers). The likelihood contours with $1\sigma$ and $2\sigma$
confidence levels for each $f(R)$ model are  presented in Fig.~1 and
2, in which the left and right panels respectively illustrate the
results on the ($a, b$) and ($\Omega_{\rm m}, b$) planes.

\subsection{$f_1(R)$ model: $f_1(R) = R - \frac{a}{R^b}$}

For the individual standard ruler data (QSO+Cluster), the best-fit
values of the parameters in the $f_1(R)$ model are
$(a,b)=(16.321,0.715)$ or $(\Omega_{\rm m},b)=(0.417,0.715)$. For
comparison, fitting results from BAO and CMB are also shown in
Fig.~1. We remark that, although the QSO+Cluster data can not
tightly constrain the model parameters, the degeneracy between
$\Omega_m$ and $b$ obtained from these individual standard rulers is
different from the statistical standard rulers (BAO and CMB).
Therefore, the quasar data, due to the wide redshift range of $0.46
< z < 2.80$, have the potential to help break the degeneracy between
model parameters in $f_1(R)$ cosmology. This tendency could also be
clearly seen from the comparison between the plots obtained with
BAO+CMB and the joint angular diameter distance data of
QSO+Cluster+BAO+CMB. With the combined standard ruler data sets, the
best-fit value for the parameters are $(a,b)=(4.669^{+0.842}
_{-0.621}, 0.042^{+0.074} _{-0.061})$ or $(\Omega_{\rm
m},b)=(0.286^{+0.018} _{-0.016}, 0.046^{+0.068} _{-0.064})$ within
68.3\% confidence level, while the constraint results from BAO+CMB
are $(a, b)=(4.349^{+1.343} _{-1.062}, 0.013^{+0.116}_{-0.112})$ or
$(\Omega_{\rm m}, b)=(0.296^{+0.034}_{-0.029}, 0.010^{+0.116}
_{-0.108})$. Therefore, the currently compiled quasar data may
significantly improve the model parameters in $f_1(R)$ cosmology.

On the other hand, the deviation parameter $b$ from joint analysis
satisfies $b=0.046^{+0.068}_{-0.064}$ at 1$\sigma$, which indicates
that $\Lambda$CDM model is still included within 68.3\% confidence
level. However, in the framework of exponential $f(R)$ gravity, the
parameter $b$ capturing the deviation from the concordance
cosmological model seems to be slightly larger than 0, which
suggests that there still exists some possibility that $\Lambda$CDM
may not the best cosmological model preferred by the current
observations. In order to make a good comparison with this standard
$\Lambda$CDM cosmology, we fix $b=0$ and obtain the marginalized
1$\sigma$ uncertainties of model parameters as:
$a=4.389^{+0.160}_{-0.180}, \ \Omega_{\rm m}=0.270^{+0.031}
_{-0.029}$ with individual standard ruler data and $a=4.249\pm
0.060, \ \Omega_{\rm m}=0.293\pm 0.011$ with combined standard ruler
data. In Table III we summarize the main results of this paper,
which are compared with recent determinations of the parameters $a$
and $b$ from independent analyses of other cosmological
observations.

\begin{table*}[t]
\begin{center}
\begin{tabular}{lccc}
\hline \hline
Data & Ref. & $a$        &       $b$\\
\hline \hline
Strong lensing & \cite{f1_R-3} & $1.50^{+12.0}_{-0.52}$ & $-0.696^{+1.21}_{-0.262}$\\
Strong lensing+BAO+CMB & \cite{f1_R-3} & $3.75^{+2.33}_{-1.29}$ & $-0.065^{+0.215}_{-0.173}$\\
BAO+CMB & This paper & $4.349^{+1.343} _{-1.062}$ & $0.013^{+0.116}_{-0.112}$ \\
QSO+Cluster+BAO+CMB & This paper & $4.669^{+0.842} _{-0.621}$ & $0.042^{+0.074} _{-0.061}$\\
\hline \hline
\end{tabular}
\end{center}
\caption{ Summary of the best-fit values for $a$ and $b$ in $f_1(R)$
model obtained from different observations.}
\end{table*}

\begin{table*}[t]
    \begin{center}
        \begin{tabular}{lccc}
            \hline \hline
            Data & Ref. & $a$        &       $b$\\
            \hline \hline
            BAO+CMB & This paper & $4.128^{+0.421}_{-0.481}$ & $-0.061^{+0.350}_{-0.343}$\\

            SNe Ia (Union 2.1)+BAO+CMB & This paper & $3.868^{+0.080} _{-0.040}$ & $-0.247^{+0.119} _{-0.144}$\\

            QSO+Cluster+BAO+CMB & This paper & $4.048^{+0.220}_{-0.281}$ & $-0.121^{+0.154}_{-0.189}$\\
            \hline \hline
        \end{tabular}
    \end{center}
    \caption{Summary of the best-fit values for $\Omega_m$ and $b$ in
        $f_2(R)$ model obtained from different observations.  }
\end{table*}

\subsection{$f_2(R)$ model: $f(R) = R - \frac{aR}{R+ab}$}

Working on the $f_2(R)$ model, we find that the best fits happen at
$(a, b)=(3.046, -0.718)$ or $(\Omega_{\rm m}, b)=(0.340, -0.718)$
with individual standard ruler data (QSO+Cluster). By marginalizing
over the parameter $a$, we derive the marginalized 1$\sigma$
constraint on the matter density parameter $\Omega_{\rm m}=0.340$,
which is well consistent with the result given by recent
\textit{Planck} first data release. The constraining power of the
individual standard rulers in breaking degeneracy between model
parameters is more obvious, as can be seen from the the marginalized
1$\sigma$ and 2$\sigma$ contours of each parameter in Fig.~2. By
fitting the $f_2(R)$ model model to QSO+Cluster+BAO+CMB, we obtain
$(a, b)=(4.048^{+0.261} _{-0.281}, -0.121^{+0.157} _{-0.189})$ or
$(\Omega_{\rm m}, b)=(0.287^{+0.017}_{-0.016},
-0.125^{+0.160}_{-0.196})$ at the $68.3\%$ confidence level.
Compared with the case in the $f_1(R)$ model, the largest difference
happens on the constraint of $b$: the deviation from $\Lambda$CDM
tends to be slightly smaller than 0, although the concordance
cosmological scenario is still included within 68.3\% confidence
level. In order to compare our fits with the results obtained using
the standard candles providing the information of luminosity
distance, we also use the latest Union2.1 compilation consisting of
580 SN Ia data \cite{SNDATA} to place constraint on the $f_2(R)$
model, which are specifically presented in Table IV. One can clearly
see that, due to the wider redshift range of the quasars data
($0.46<z<2.80$) compared with SN Ia ($0.015<z<1,41$), the current
standard ruler data make a good improvement on the constraints of
the $f_2(R)$ model parameters. The recent determinations of the
parameters $a$ and $b$ from other independent cosmological
observations are also listed in Table IV.

\begin{table}[htp]
\centering

{
\begin{tabular}{c  c  c c  c c c} \hline\hline
                $Model$  & $k$ & $\chi^2_{min}$ & $AIC$ & $\Delta AIC$ & $BIC$ & $\Delta
                BIC$\\ \cline{1-7}

                $\Lambda CDM$ & $1$ & 384.86 & $386.86$ & $0$ & $389.88$ & $0$ \\
                $f_1(R)$ & $2$ & 383.87 & $387.87$ & $1.01$ & $393.92$ & $4.04$ \\
                $f_2(R)$ & $2$ & 383.57 & $387.57$ & $0.71$ & $393.62$ & $3.74$ \\
                \cline{1-7}
                \hline \hline
        \end{tabular} \label{tableIC}}
\caption{ Summary of the minimum $\chi^2$ and the information
criteria for the two $f(R)$ models, obtained from the combinations
of standard rulers: CMB+BAO+QSO+Cluster. Corresponding results for
the $\Lambda$CDM are also added for comparison.}
\end{table}

\subsection{Model selection}

Based on a likelihood method, one may employ the information
criteria (IC) to assess different models. In order to decide which
$f(R)$ model is favored by the observational data, we perform model
comparison statistics by using the Akaike Information Criterion
(AIC) \cite{AIC} and the Bayesian Information Criterion (BIC)
\cite{BIC}. The expressions of the two information criteria are
respectively given by
\begin{equation}
AIC=-2\mathrm{ln}\mathcal{L}_{max}+2k=\chi^2_{min}+2k
\end{equation}
and
\begin{equation}
BIC=-2\mathrm{ln}\mathcal{L}_{max}+k\mathrm{ln}N=\chi^2_{min}+k\mathrm{ln}N
\end{equation}
where $\chi^2_{min}$ is the minimum $\chi^2$ value, while $k$ and
$N$ represent the total number of model parameters and data points.
As can be clearly seen from the two information criteria, models
that give a good fit with fewer parameters will be more favored by
observations. The application of the BIC and AIC in a cosmological
context can be found in the previous study
\citep{cao2012testing,pan2016constraints}.

In this analysis, we use the the combined standard ruler data
(quasars, galaxy clusters, BAO and CMB) for the purpose of model
selection. Table~V lists the value of the minimum $\chi^2$, AIC, BIC
and the corresponding IC difference of each model ($f_1(R)$,
$f_2(R)$ and $\Lambda$CDM). Note that the cosmological constant
model has the lowest value of IC and the value of $\Delta$IC is
measured with respect to this model. One will find that, given the
current individual standard ruler data combined with the BAO and CMB
data, the information criteria indicate that the cosmological
constant model is still the best one, since both the AIC and BIC
values it yields are the smallest. Concerning the ranking of the two
$f(R)$ models, AIC and BIC criteria give very similar conclusions,
which indicate that $f_2(R)$ is slight favored, while the $f_1(R)$
model gets the smallest support from the current observations. More
specifically, considering the fact that a difference in BIC
($\Delta$BIC) of 2 is positive evidence against the model with
higher BIC \cite{BIC}, our findings show that $f(R)$ cosmologies
with more free parameters are penalized by the BIC criterion. This
conclusion is in accordance with the previous results derived from
other cosmological probes \citep{cao2012testing,Zheng17ultra}.

In order to gain more insights into the above findings, in the
following analysis we will turn to other sensitive and robust
diagnostics to illustrate the dynamic behavior of different
cosmologies. The corresponding results are obtained based on the
best fits from the joint analysis of standard ruler data.

\section{The model diagnostics}

In the framework of a specific cosmological model, the Hubble
parameter $H$ and the deceleration parameter $q$ respectively
express as
\begin{equation}
\label{H_q} H=\frac{\dot{a}}{a},
q=-\frac{\ddot{a}}{aH^2}=-\frac{a\ddot{a}}{\dot{a} ^2},
\end{equation}
where $a$ is the scale factor. These quantities were first propose
to test both the evolution of cosmology (i.e., the determination of
transition redshift (deceleration/acceleration), which has been
proved to provide an efficient way for constraining cosmological
models \citep{chen2010using}), and then extensively applied to the
investigation of the dynamical properties of dark energy (i.e., the
possible interaction between cosmic dark sectors
\citep{cao2010testing,cao2011interaction}). In this section, we will
perform two diagnostics analysis based on the above two quantities,
i.e., the $Om(z)$ diagnositc and the statefinder diagnostic, to
discuss the possibility of discriminating the three cosmological
models ($f_1(R)$, $f_2(R)$ and $\Lambda$CDM).

\begin{figure*}
    \centering
    \includegraphics[width=8cm,height=7.2cm]{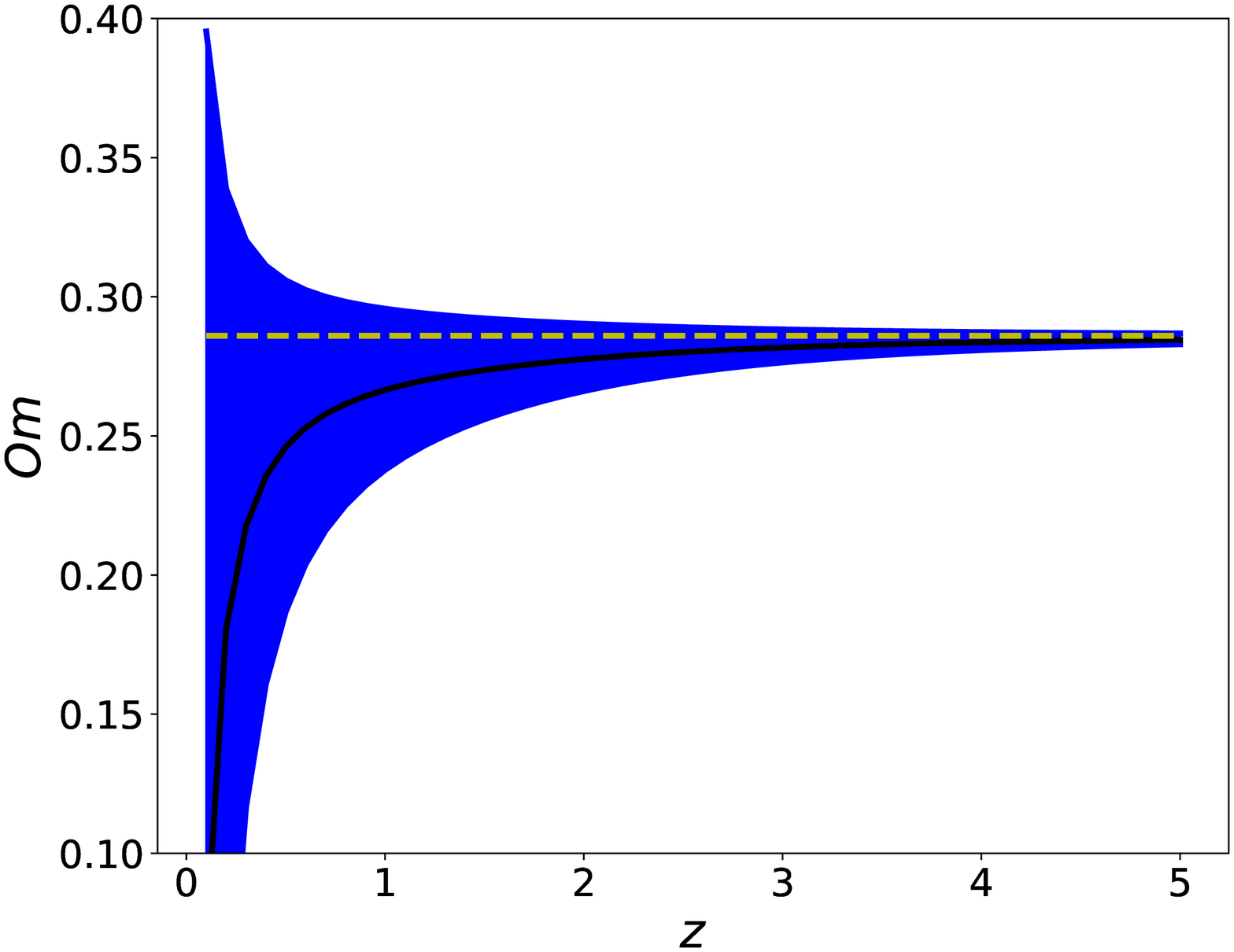}\includegraphics[width=8cm,height=7.2cm]{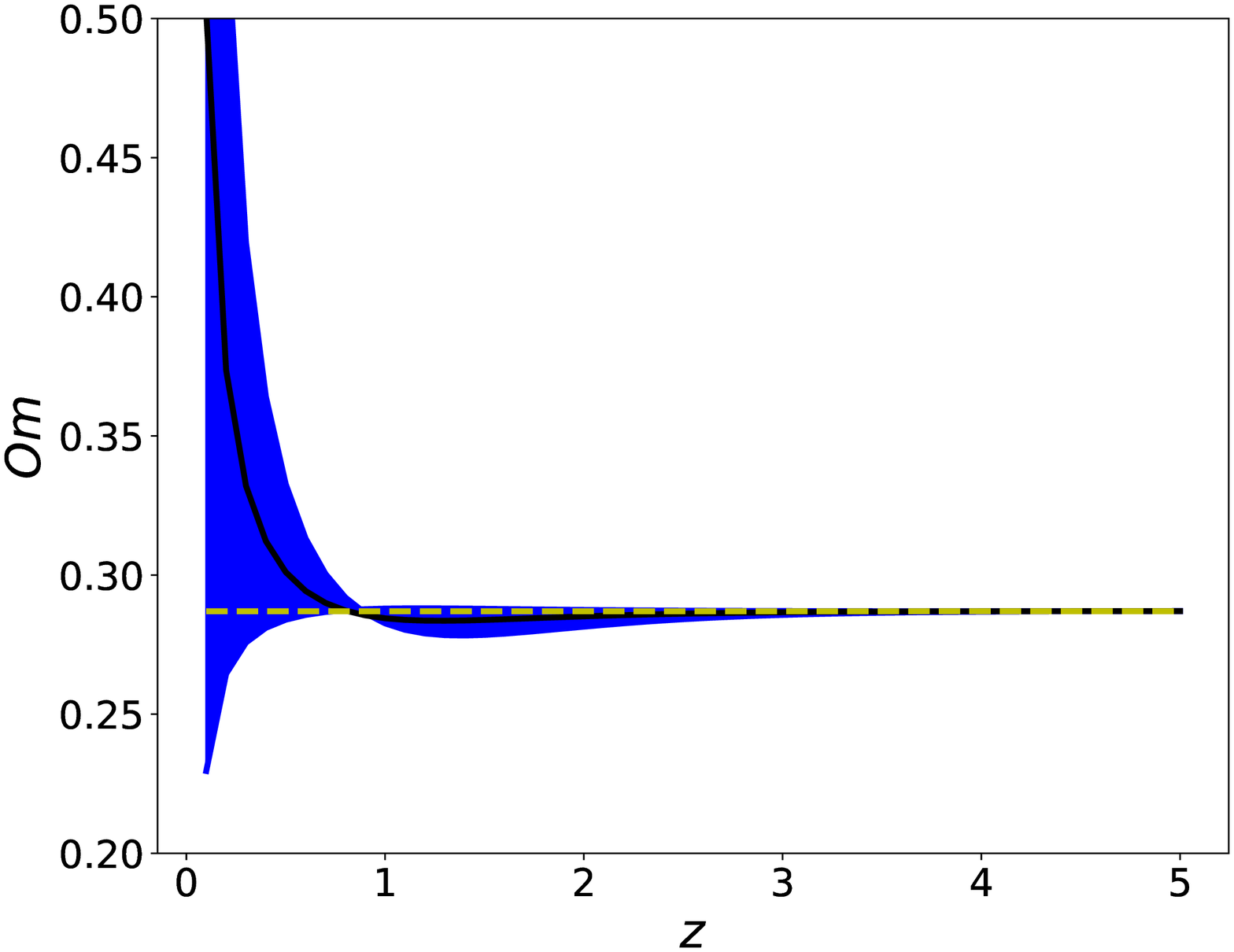}
\caption{ The evolution of $Om(z)$ versus the redshift $z$ for
$f_1(R)$ model (left panel) and $f_2(R)$ model (right panel) from
the combined standard ruler data (black solid line), with the
1$\sigma$ uncertainty denoted by blue shades. The standard
$\Lambda$CDM model (yellow dashed line) is also added for
comparison.  }\label{fig:Omz}
\end{figure*}

\subsection{$Om(z)$ diagnostic}

The expansion rates at different redshifts, or the Hubble parameters
$H(z)$, opened a new chapter in using the so called $Om(z)$
diagnostics to discriminate different cosmological models as well as
$\Lambda$CDM model \cite{Om_diagnostic}. In this method, a new
diagnostics is defined as
\begin{equation}
\label{Om_z} Om(z)=\frac{E^2(z)-1}{(1+z)^3-1}
\end{equation}
where $E(z)=H(z)/H_0$ is the dimensionless expansion rate.
Neglecting the radiation component at low redshift, the Friedmann
equation in the $\Lambda$CDM model is $H(z)^2=H_0 ^2[\Omega_{\rm
m}(1+z)^3+1-\Omega_{\rm m}]$ and the $Om(z)$ parameter should be
equal exactly to the present value of matter density
$Om(z)=\Omega_m$, if the cosmological model of our universe is
exactly the $\Lambda$CDM model. Therefore, such diagnostic can be
used as a cosmological probe to directly illustrate the difference
between $\Lambda$CDM and other cosmological models.

Fig.~3 shows the $Om(z)$ parameter as a function of redshift for the
two $f(R)$ models, with the best-fitted value as well as the
1$\sigma$ uncertainties derived from the combined standard ruler
data. It is obvious that the $Om(z)$ curves of the two $f(R)$ models
will coincide with $\Lambda$CDM at high redshifts ($z>3$), which
indicates that the $Om(z)$ for the $f(R)$ models cannot be
distinguished $\Lambda$CDM at early universe. The $f(R)$ cosmology
begins to deviate from the $\Lambda$CDM at the redshift interval of
$1.5<z<3$, although we still cannot distinguish the two $f(R)$
models at this epoch. At redshifts $z<1$, the two $f(R)$ models
begin to deviate with each other: for $f_1(R)$ model, the value of
$Om(z)$ is always smaller than that in $\Lambda$CDM model, while the
deviation between $f_1(R)$ and $\Lambda$CDM gradually increase,
which implies a smaller $Om(z)$ in the future; for $f_2(R)$ model,
the other cosmological candidate proposed without introducing dark
energy in the Universe, the value of $Om(z)$ is also smaller than
$\Omega_{\rm m}$ when the deviation from $\Lambda$CDM model takes
place, which will increase and exceed $\Omega_{\rm m}$ at lower
redshifts.

\subsection{Statefinder diagnostic}

As can be clearly seen from Eq.~(\ref{H_q}), the Hubble parameter
$H$ and the deceleration parameter $q$ are respectively related to
$\dot{a}$ and $\ddot{a}$. Considering that $H$ and $q$ cannot
effectively differentiate between different cosmological models, a
joint analysis still requires a new parameter related to the higher
order of time derivatives of $a$. In the so-called statefinder
diagnostic \cite{SF}, the statefinder pair $\{r,s\}$ is defined as
\begin{equation}
r=\frac{\dddot{a}}{aH^3}, \quad s=\frac{r-1}{3(q-1/2)}
\end{equation}
where the deceleration parameter can be derived from Eq.~(\ref{H_q})
as
\begin{equation}
\label{q_z} q(z)=\frac{E^\prime (z)}{E(z)}(1+z)-1.
\end{equation}
where $E^\prime (z) \equiv \mathrm{d} E(z)/\mathrm{d} z$. Then the
statefinder pair $\{ r,s\}$ can be expressed as
\begin{eqnarray}
\nonumber r(z)&=&1-2\frac{E^\prime
(z)}{E(z)}(1+z)+\left[\frac{E^{\prime
\prime}(z)}{E(z)}+\left(\frac{E^\prime
(z)}{E(z)}\right)^2\right](1+z)^2
\\
\label{r_z} &=&q(z)(1+2q(z))+q^\prime (z)(1+z)
\\
\label{s_z} s(z)&=&\frac{r(z)-1}{3(q(z)-1/2)}
\end{eqnarray}
where $q^\prime (z) \equiv \mathrm{d} q(z)/\mathrm{d} z$. Applying
the best-fitted value as well as the 1$\sigma$ uncertainties derived
from the combined standard ruler data for each $f(R)$ model, we
figure out the evolution of statefinder pair $(r, s)$ and the
deceleration parameter $q$, and show the trajectories of the two
$f(R)$ models in the $r-s$ and $r-q$ plane in Fig.~4 and 5. It is
worth mentioning that, the statefinder parameters $r$ and $s$ for
$\Lambda$CDM model are constants, $(r, s)=(1, 0)$ \cite{SF_r_s_1_0}.

The evolution trajectories in the $r-q$ plane for different
cosmological models are shown in Fig.~4. One may note that the
deceleration parameter $q$ evolves similarly for the two $f(R)$
models and $\Lambda$CDM model, while the statefinder parameter $r$
exhibits obvious fluctuation, which indicates that the parameter $r$
is more suitable to discriminate different cosmological models. As
can be clearly seen from Fig.~4, the curve of each cosmological
model originates from the same point $(r, q)=(1, 0.5)$, evolves
along different trajectory, and finally converges on the same point
$(r, q)=(1, -1)$. For $f_1(R)$, the deviation from $\Lambda$CDM is
not as obvious as that between $f_2(R)$ and $\Lambda$CDM, while the
trajectory of $f_1(R)$ converges with $\Lambda$CDM earlier than
$f_2(R)$. At the present epoch, the parameters $(r, q)$ for $f_1(R)$
and $\Lambda$CDM cannot be distinguished from each other, while the
value of $r$ for $f_2(R)$ is lager than that of the other two
models. More interestingly, it is noteworthy that for $f_2(R)$
cosmology, we observe the signature flip from negative to positive
in the value of $r$ in the early universe, which indicates that the
$f_2(R)$ model is quite different from $f_1(R)$ model and
$\Lambda$CDM model at early times.

The evolution of statefinder pair ($r, s$) for different
cosmological models are plotted in Fig.~5, in which the yellow
square at $(r, s) =(0, 1)$ indicates the statefinder of $\Lambda$CDM
model. It is apparent that the statefinder pair of $f_2(R)$ evolves
quite different from that in the framework of $f_1(R)$CDM model.
More importantly, as is shown in Fig.~5, the current value of $(r,
q)$ for $f_1(R)$ is well consistent with $\Lambda$CDM, while the
corresponding value for $f_2(R)$ significantly deviates from both of
them at the present epoch. Therefore, although the statefinder pair
$(r, q)$ cannot distinguish $f_1(R)$ model and $\Lambda$CDM model
from each other at the present time, it is able to distinguish the
$f_2(R)$ from both $f1$CDM and $\Lambda$CDM. However, the $f(R)$
cosmologies are not distinguishable and can not be distinguished
from $\Lambda$CDM in the near future.

As a final comment, one should note that the combination of standard
ruler data implies that, the two modified gravity models discussed
in this analysis are still practically indistinguishable from
$\Lambda$CDM. Such a tendency is more obvious when the 1$\sigma$
uncertainty of the model parameters is taken into consideration (See
Figs.~3-5). Therefore, the above conclusions still need to be
checked by future high-precision VLBI observations of the compact
structure in radio quasars \cite{Pushkarev15,cao2017multif}, which
also highlights the importance of different cosmological standard
rulers to provide additional observational fits on alternative
candidate gravity theories.

\begin{figure*}
    \centering
    \includegraphics[width=8cm,height=7.2cm]{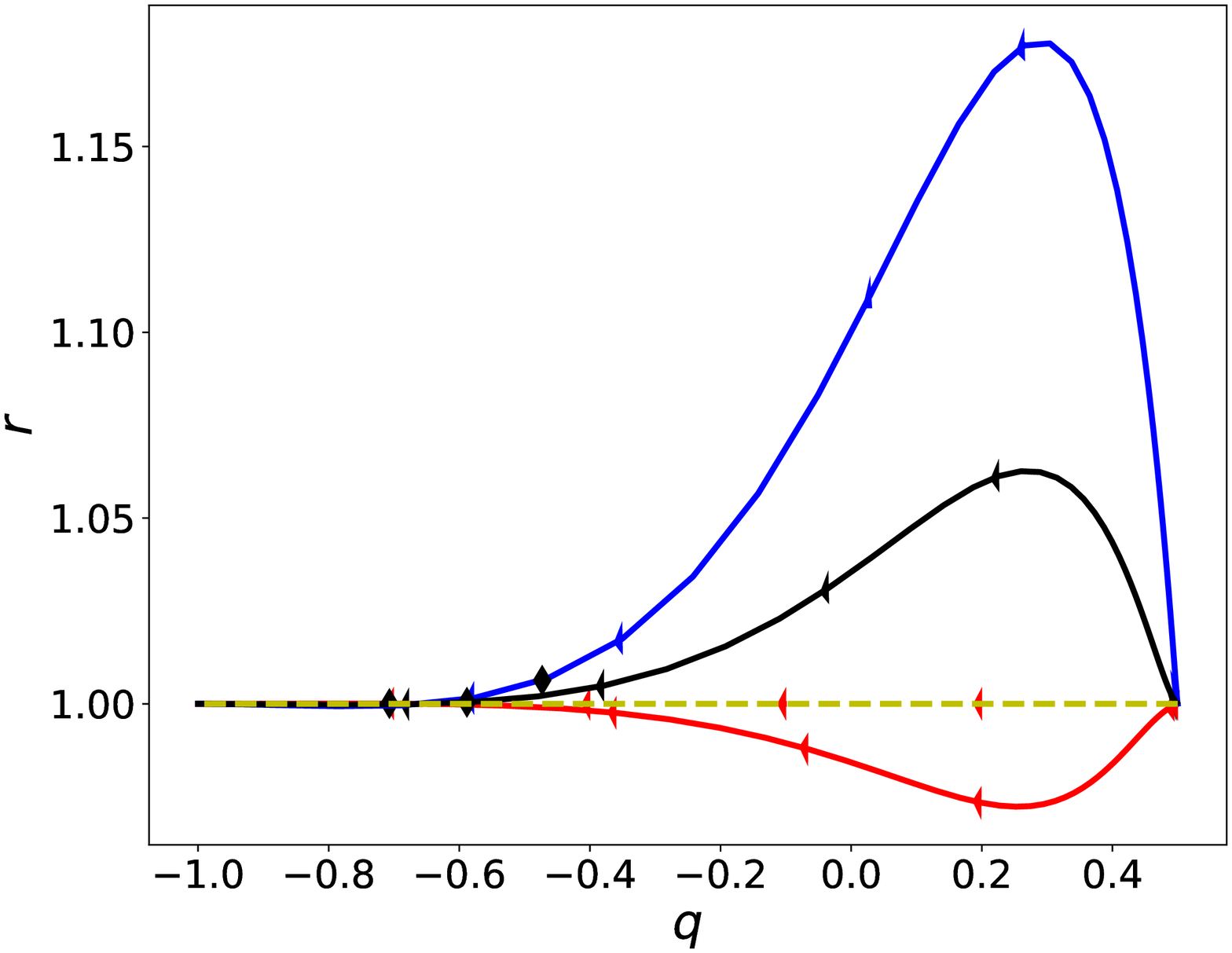}\includegraphics[width=8cm,height=7.2cm]{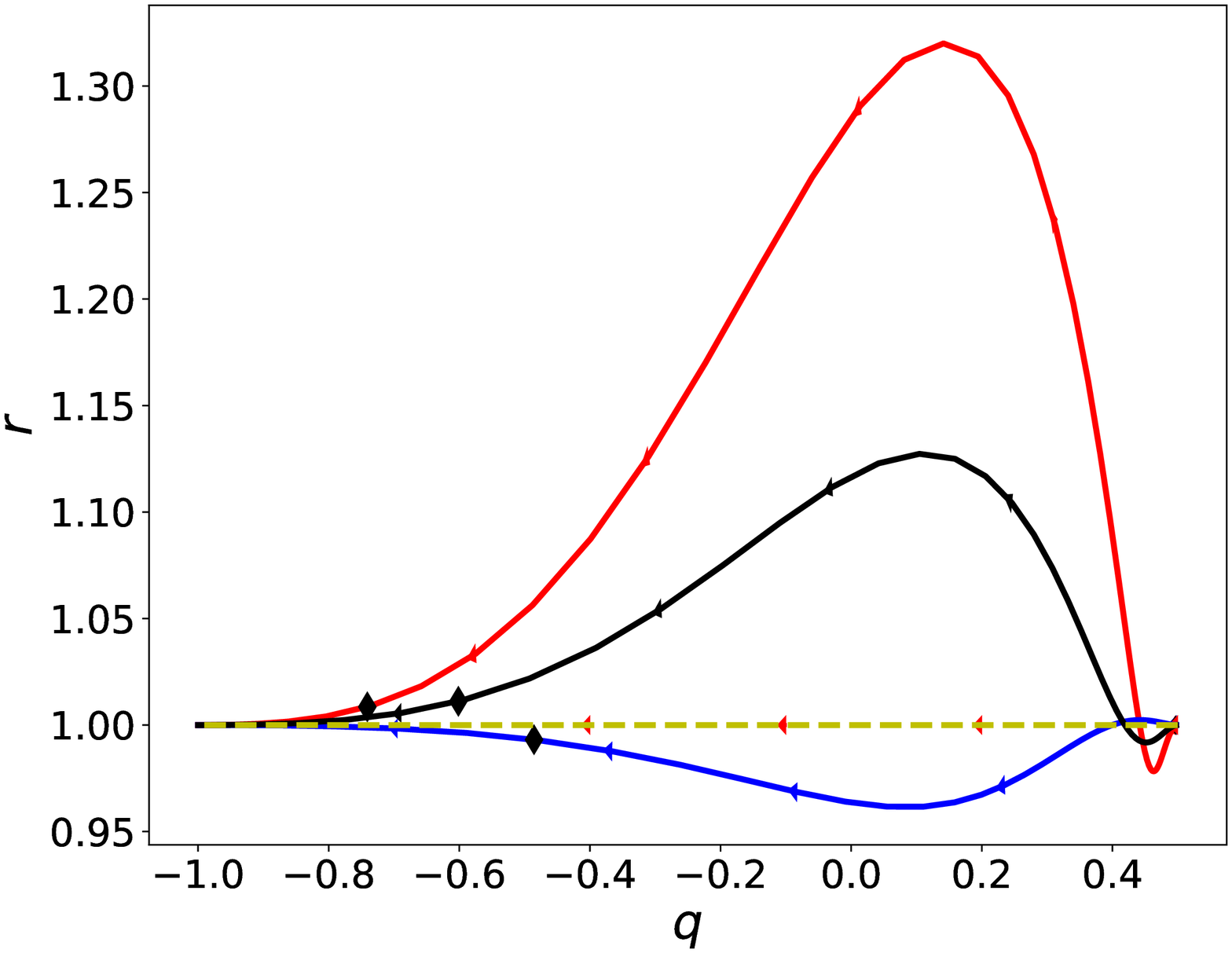}
\caption{ The evolution of $(r, q)$ for $f_1(R)$ model and $f_2(R)$
model (black line), with the 1$\sigma$ uncertainty denoted by red
and blue lines. The standard $\Lambda$CDM model (yellow dashed line)
is also added for comparison. The diamond point and the arrows on
each curve respectively denote the current value and the evolution
direction of $(r, q)$ for each cosmological model. }\label{fig:qr}
\end{figure*}

\begin{figure*}
    \centering
    \includegraphics[width=8cm,height=7.2cm]{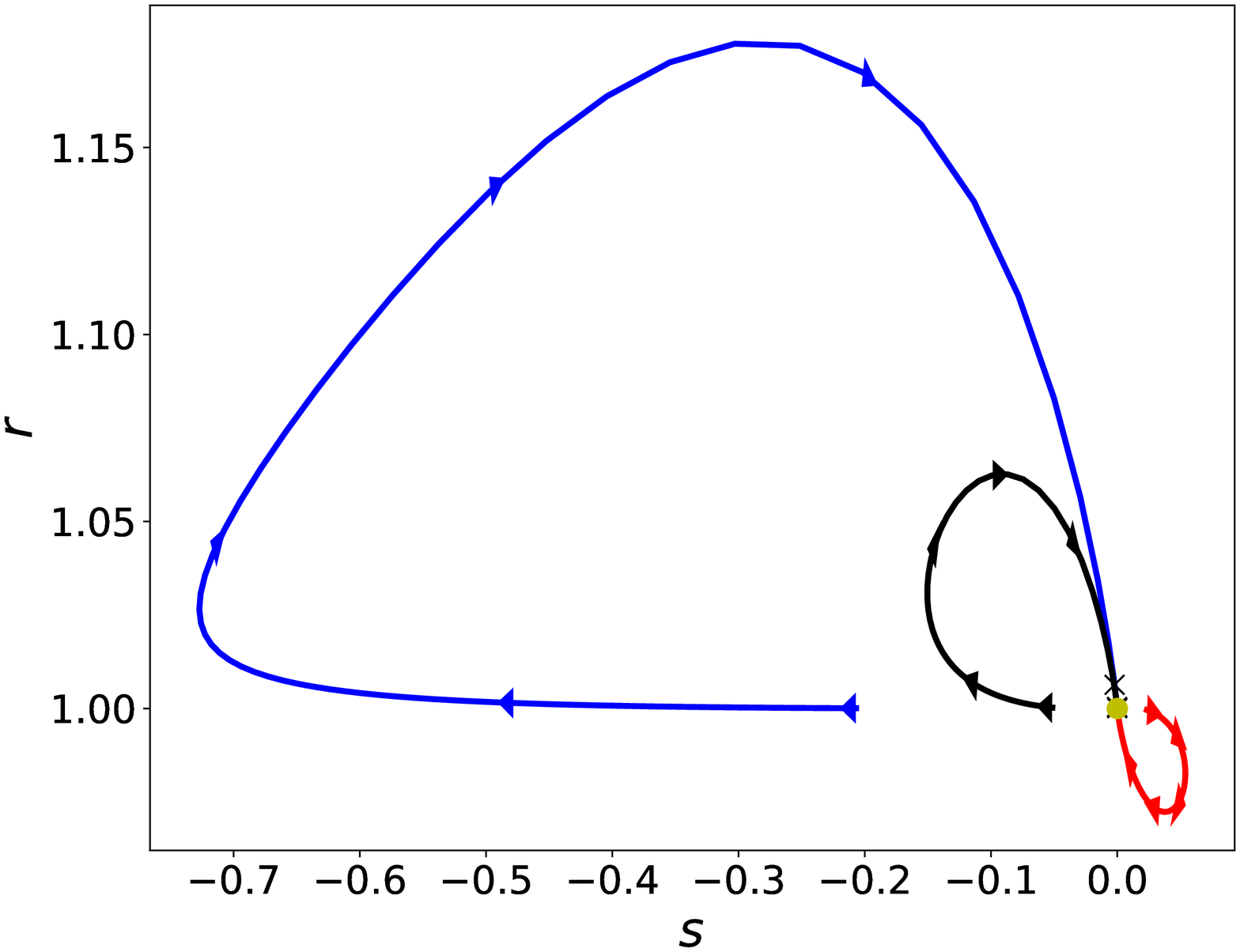}\includegraphics[width=8cm,height=7.2cm]{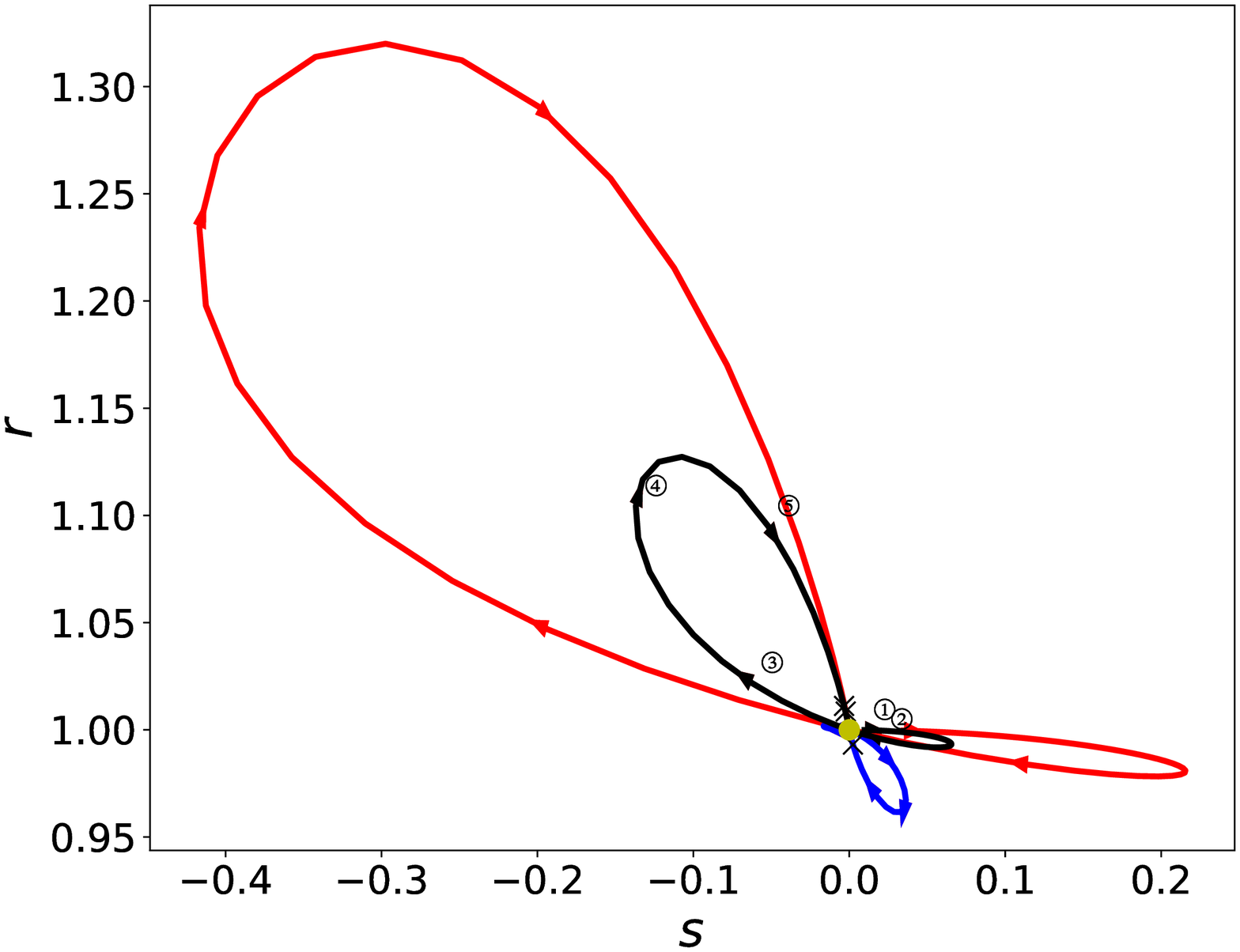}
\caption{ The same as Fig.~4, but for the evolution of $(r, s)$. The
black cross and the arrows on each curve respectively denote the
current value and the evolution direction of $(r, s)$ for each
cosmological model. The yellow square at $(r, s)=(1, 0)$ represents
$\Lambda$CDM model. For $f_2(R)$, in order to represent the
evolution trajectory more clearly, the circled sequence numbers are
marked on the curve. }\label{fig:sr}
\end{figure*}

\section{Conclusions and discussions}

As an important candidate gravity theory alternative to dark energy,
a class of $f(R)$ modified gravity, which introduces a perturbation
of the Ricci scalar $R$ in the Einstein-Hilbert action, has been
extensively applied to cosmology to explain the acceleration of the
universe. On the other hand, recent observations of various
cosmological standard rulers acting as distance indicators in the
Universe, have provided a lot of information concerning angular
diameter distance $D_A$. In this paper, we focused on the
recently-released VLBI observations of the compact structure in
intermediate-luminosity quasars combined with the
angular-diameter-distance measurements from galaxy clusters, which
consists of 145 data points performing as individual cosmological
standard rulers in the redshift range $0.023\le z\le 2.80$, to
investigate observational constraints on two viable models in $f(R)$
theories within the Palatini formalism: $f_1(R)=R-\frac{a}{R^b}$ and
$f_2(R)=R-\frac{aR}{R+ab}$. Here we summarize our main conclusions
in more detail.

\begin{itemize}

\item
In the framework of $f(R)$ gravity in Palatini approach, although
the QSO+Cluster data can not tightly constrain the model parameters,
the degeneracy between $\Omega_m$ and $b$ obtained from these
individual standard rulers is different from the statistical
standard rulers (BAO+CMB). Therefore, the quasar data have the
potential to help break the degeneracy between model parameters in
$f(R)$ cosmology. More specifically, one can clearly see that, due
to the wider redshift range of the quasars data ($0.46<z<2.80$)
compared with SN Ia ($0.015<z<1.41$), the current standard ruler
data make a good improvement on the constraints of the $f_2(R)$
model parameters.

\item
From the joint analysis of combined standard ruler data, the matter
density parameter $\Omega_{\rm m}$ derived from the combined
standard ruler data in the $f_1(R)$ and $f_2(R)$ model are both
close to 0.30, which is consistent with that from other independent
cosmological observations. Moreover, with the assumption of a flat
universe and given the current quality of the observational data,
the $\Lambda$CDM model is still included within 68.3\% confidence
level and there is no reason to prefer any more complex model. Given
the current individual standard ruler data combined with the BAO and
CMB data, the information criteria imply that the cosmological
constant model is still the best one, while the $f_1(R)$ model gets
the smallest support from the current observations.

\item
Deviation from $\Lambda$CDM cosmology is also detected in the
obtained confidence level in our analysis. More specifically, in the
framework of exponential $f_1(R)$ gravity, the deviation parameter
$b$ denoting the difference between $f(R)$ gravity and the
concordance $\Lambda$CDM model seems to be slightly larger than 0,
while for $f_2(R)$ gravity, the largest difference happens on the
constraint of $b$, i.e., the deviation from $\Lambda$CDM tends to be
slightly smaller than 0. Therefore, there still exists some
possibility that $\Lambda$CDM may not the best cosmological model
preferred by the current observations. However, this conclusion
still needs to be checked and confirmed by future more accurate
observational data.

\item
Applying the best-fits from combined standard ruler data to two
model diagnostics, $Om(z)$ and statefinder, we have applied two
model diagnostics to differentiate the dynamical behavior of the
$f(R)$ models. The results from the $Om(z)$ diagnostic show that the
$Om(z)$ for the $f(R)$ models cannot be distinguished $\Lambda$CDM
at early universe. As the redshift decreases, the $f(R)$ cosmology
begins to deviate from the $\Lambda$CDM at the redshift interval of
$1.5<z<3$. On the other hand, the statefinder diagnostic indicates
that in the early universe $f_2(R)$ evolves quite different from
that in the framework of $f_1(R)$ model, both of which will finally
evolve to the same state. At the present time, $f_1(R)$ model is
well consistent with $\Lambda$CDM model, while $f_2(R)$ model still
significantly deviates from both $f_1(R)$ and $\Lambda$CDM model.
However, when the 1$\sigma$ uncertainty of the model parameters is
taken into consideration, these two modified gravity models are
still practically indistinguishable from $\Lambda$CDM.

\item Given the redshift coverage
of high-redshift quasars and low-redshift clusters, the combination
of these two astrophysical probes could provide a relatively
complete source of angular diameter distances. From the perspective
of observations, the recently released quasar data propose a new way
to probe the cosmology. Furthermore, the quasar data used in this
paper were observed at single frequency, we expect the future
high-precision observations derived at multi frequencies
\cite{Pushkarev15,cao2017multif} to provide more information of
other classes of modified gravity theories. From the theoretical
point of view, besides the two $f(R)$ models which we have already
extensively investigated in this paper, some other typical examples
are Lorentz violating theories \cite{1475-7516-2007-08-010}, ghost
condensation \cite{1126-6708-2004-05-074}, and tensor-vector-scalar
theory of gravity \cite{PhysRevD.71.069901}.

\end{itemize}

\section{Acknowledgments}

This work was supported by National Key R\&D Program of China No.
2017YFA0402600, the National Basic Science Program (Project 973) of
China under (Grant No. 2014CB845800), the National Natural Science
Foundation of China under Grants Nos. 11503001, 11633001 and
11373014, the Strategic Priority Research Program of the Chinese
Academy of Sciences, Grant No. XDB23000000, the Interdiscipline
Research Funds of Beijing Normal University, and the Opening Project
of Key Laboratory of Computational Astrophysics, National
Astronomical Observatories, Chinese Academy of Sciences. J.-Z. Qi
was supported by China Postdoctoral Science Foundation under Grant
No. 2017M620661. M. Biesiada was supported by Foreign Talent
Introducing Project and Special Fund Support of Foreign Knowledge
Introducing Project in China.

%
%

\end{document}